\documentclass[11pt]{article}

\usepackage{geometry}
\usepackage{multicol, caption}

\usepackage{graphicx}
\usepackage{amsthm}
\usepackage{booktabs}

\usepackage{amsmath,amssymb} 
\usepackage[utf8]{inputenc}
\usepackage{amsmath}
\usepackage{amssymb}  
\usepackage{pifont}
\usepackage{pstricks}
\usepackage{amsfonts}
\usepackage{graphicx}
\usepackage{amsmath}
\usepackage{authblk}
\usepackage{pstricks} 
\usepackage{pstricks-add}
\usepackage{framed}
\usepackage{bm}
\usepackage{amsfonts}
\usepackage{amsmath}
\usepackage[T1]{fontenc}
\usepackage{pifont}
\usepackage{caption}
\usepackage{placeins}
\usepackage{pstricks}
\usepackage{pdfpages}
\usepackage{amsfonts}
\usepackage{graphicx}
\usepackage{placeins}
\usepackage{amsmath}
\usepackage{authblk}
\usepackage{pstricks}
\usepackage{pstricks-add}
\usepackage{framed}
\usepackage{bm}

\newenvironment{Figure}
  {\par\medskip\noindent\minipage{\linewidth}}
  {\endminipage\par\medskip}

\title{Calculation of apsidal precession via perturbation theory}
\author{L.~Barbieri, F.~Talamucci}
\affil{DIMAI -- Dipartimento di Matematica e Informatica \\
University of Florence, Italy\\
e-mail: federico.talamucci@unifi.it}
\date{}

\begin{document}
\bibliographystyle{plain}



\noindent
{\LARGE\bf Calculation of apsidal precession via perturbation theory}

\vspace{.7truecm}

\noindent
{\large L.~Barbieri, F.~Talamucci$^{1,*}$}

\vspace{.5truecm}

\noindent
\footnotesize{$\,^{1}$DIMAI -- Dipartimento di Matematica e Informatica, 
University of Florence, Italy}

\date{}


\vspace{.5truecm}

\noindent
{\large\bf Email address}

\vspace{.5truecm}

\noindent
federico.talamucci@unifi.it (F.~Talamucci)

\vspace{.5truecm}
$\,^{*}$Corresponding author

\vspace{.5truecm}

\noindent
{\large\bf Abstract}

\vspace{.3truecm}

\noindent
The calculus of apsidal precession frequencies of the planets is developed by means of a perturbation thecnique. A model of concentric rings ({\it ring model}), suitable for improving calculations, is introduced.
Conclusive remarks concerning a comparison between the theoretical, the calculated and the observed data of the precession frequencies are performed.

\vspace{.5truecm}

\noindent
{\large\bf Keywords}

\vspace{.3truecm}

\noindent
Orbital Mechanics, apsidal precession frequencies, ring model

\vspace{.5truecm}

\begin{multicols}{2}

\section{Introduction}

\noindent
The solar system is a gravitationally bound system encompassing the Sun, the planets and many other celestial bodies.
As it is known, apsidal precession consists in the rotation of a planet apsidal line, which is the line passing through aphelion and perihelion. The precession of each planet is caused by the gravity effects of celestial bodies (in particular other planets) or by a relativistic effect. We will consider only classical effects and we will shortly refer to the relativistic ones only at the end of the work.

\noindent
We will focus on the simplified system formed by the Sun and the planets, disregarding the rest of the bodies. The Keplerian model is able to reproduce the revolution motion whenever the apsidal line is fixed: the starting point consists in showing how the apsidal line precedes and the eccentricity varies, if a force $\textbf{F}$, ascribable, as an example the presence of other planets, is added to the Keplerian force.

In the second part we make use of the ``ring model'' in order to calculate the apsidal precession frequencies. 
The basic assumption of such model is that the precession period of every single planet is much greater than any other revolution period. In this way, the model deals with the planets alterating the revolution motion as concentric rings centered in the Sun with a uniform mass distribution.
In particular, by means of suitable geometrical properties, it is possible to consider orbits as coplanar and circular. Such approximation let us enter the central force formalism and the perturbation theory is applied directly to the orbit equation. 

\noindent
Secondly, in order to calculate the frequency of precession, we make use of the perturbation theory performed in \cite{Pollard}. We will calculate the precession frequencies of each planet and we will comment the differences between theoretical frequencies, calculated frequencies and the observed data.

\section{Apsidal precession}

\noindent
According to \cite{Precession2}, we will discuss the effect of a force $\textbf{F}$ added to the Keplerian force. We consider a mass $m$ particel subjected to a total force which is the sum of a force $\textbf{F}$ and the Keplerian force:
$$
\textbf{F}_{tot}=-\frac{k}{r^2}\hat{\textbf{r}}+\textbf{F}.
$$
In the Keplerian motion $\textbf{F}={\bf 0}$ the Lenz vector $\textbf{A}=m \textbf{p}\times {\bf L}-mk\hat{\textbf{r}}$ (where $\textbf{p}$ is the momentum and ${\bf L}=\textbf{r}\times \textbf{p}$ is the angular momentum) is a constant of motion whose direction defines the apsidal line and the magnitude is proportional to the eccentricity. Because of the perturbation introduced by the force $\textbf{F}$ the Lenz vector $\textbf{A}$ is no longer a constant of motion and it changes in direction and in magnitude. The change in direction causes the line precession and the change in magnitude produces variation in time of the eccentricity.
An important property of the apsidal precession motion can be deduced by writing the precession frequency 
\begin{equation}
\label{freq}
\nu=\frac{\textbf{A} \times \dot{\textbf{A}}}{||\textbf{A}||^2}
\end{equation}
where $\dot{\textbf{A}}=m(2(\dot{\textbf{r}}\cdot \textbf{F})\textbf{r} - (\textbf{r} \cdot \dot{\textbf{r}})\textbf{F} -(\textbf{r} \cdot \textbf{F})\dot{\textbf{r}})$.
As we can see, (\ref{freq}) is linear with respect to $\textbf{F}$, hence if the perturbation $\textbf{F}$ is the sum of two forces, then the precession frequency due to $\textbf{F}$ is the sum of the two frequencies of the singular contribution due to the first and the second force. We are going now to analyse the remarkable case  where the perturbative force is 
$$
\textbf{F}(r)= \frac{H}{r^3}\dfrac{\hat{\textbf{r}}}{r}
$$
where $H$ is a negative number. In the next Section we will find that this is a for the gravitational perturbation on the revolution motion of planets.
The effective potential energy is:
\begin{equation}
\label{poteff}
U_{eff}=-\frac{k}{r}+\frac{||L||^2+m H}{2m r^2}
\end{equation} 
In most significant cases the ratio $\frac{m H}{||\textbf{L}||^2}$ is much smaller then 1. As we can see from (\ref{poteff}), the effective potential energy is the same as the Keplerian case, so that the orbit is limited between two values of $r$. The orbit equation is
$$
\frac{d^2 \rho}{d\theta^2} + \biggl(1+\frac{m H}{||\textbf{L}||^2}\biggl)\rho=\frac{d^2 \rho}{d\theta^2} + \Omega ^2\rho=\frac{m k}{||\textbf{L}||^2}
$$
corresponding to the equation of a driven harmonic oscillator with a driving force given by $\frac{m k}{||\textbf{L}||^2}$. The solution can be easily computed in terms of the inverse radius as
\begin{equation}
\label{soluzinv}
\rho=\frac{1}{r}=\frac{m k}{||\textbf{L}||^2 \Omega ^2}(1+e\cos(\Omega (\theta -\omega)))
\end{equation}
where $e$ and $\omega$ are two integration constant values. In the case of interest the factor $\Omega$ is close to $1$ and smaller than $1$: the curve (\ref{soluzinv}) is sketched in Figure 1 in the case $\Omega=0.9$. As we can see, the effect is a rotation of the Keplerian orbit in its plane.


\begin{Figure}
\centering
\includegraphics[scale=.5]{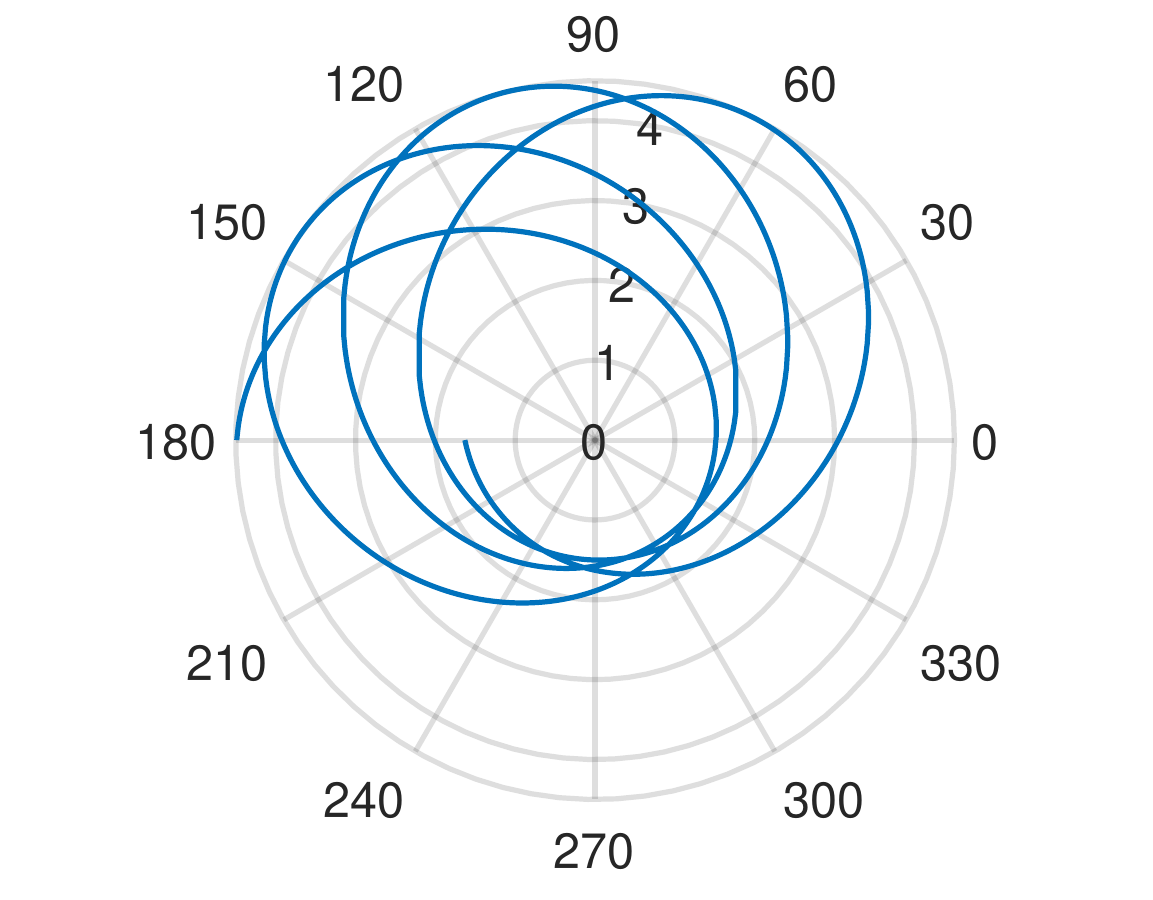}
\captionof{figure}{the orbit (\ref{soluzinv}), $\Omega=0.9$.}
\end{Figure}


\section{The ring model}
We introduce now a model based on the assumption that the precession period of every single planet is much greater than any other revolution period. As a consequence, we can study the precession motion of a specific planet by treating the rest of the planets as uniform concentric rings, centered in the Sun, whose masses are equal to the respective planetary masses. Furthermore, each radius is equal to the respective mean orbital radius. 

\noindent
On the basis of this assumption, we simplify the geometry of the problem 
by using specific properties of the solar system. By virtue of the small eccentricities, 
the rings can be approximated as  circles. On the other hand, the small inclination of the orbital planes from the ecliptic plane allow us to approximate the orbits as coplanar orbits (central force approximation).

\noindent
By employing this model we calculated the precession frequencies using perturbation techniques illustrated in Chapter 4 of \cite{Pollard} . The perturbation is caused by the gravity of the other planets (treated as  rings) which alter the Keplerian motion of the single planet. 

\subsection{Ring's gravity field}

\noindent
Consider a uniform ring of mass $m$ and radius $R$ and a polar coordinate system $(r,\theta)$ in the plane of the ring and centered in its center. The gravitational potential is
\begin{equation}
\label{gravitpot}
V(r)=-\frac{Gm}{2\pi R} \int_{0}^{2\pi} \frac{Rd\theta}{\sqrt{r^2+R^2-2rR cos \theta}}
\end{equation}
that is a complete elliptical integral of the first kind. In order to 
exhibit the solution, we have to make a distinction between the case of internal field ($r<R$) and external field ($r>R$).

\paragraph{Internal field}
Expanding the argument of (\ref{gravitpot}) in powers of $\frac{r}{R}$ and calculating the integral, one can obtain the formula (see \cite{Precession1})
$$
V(r)=-\frac{Gm}{R}\biggl(1+\sum_{n=1}^{+\infty}\biggl(\frac{(2n-1)!!}{(2n)!!}\biggl)^2 \biggl(\frac{r}{R}\biggl)^{2n}\biggl).
$$

\noindent
We can calculate the gravitational field in the case $r<R$:
\begin{eqnarray}
\nonumber
g_{int}(r)&=&\frac{Gm}{R^2}\biggl(\sum_{n=1}^{+\infty}\biggl(\frac{(2n-1)!!}{(2n)!!}\biggl)^2(2n)\biggl(\frac{r}{R}\biggl)^{2n-1}\biggl)\\
\label{gint}
&=&\sum_{n=1}^{+\infty}\alpha_{n}r^{2n-1}
\end{eqnarray}
In order to simplify the latter expression, we follow a procedure showed in \cite{PrecessionN} and based on 
a comparison technique between series. Let $r_1$ and $r_2$ be respectively the radial distance of the aphelion and the perihelion of a planet subjected to the field (\ref{gint}) and calculate the constant values  $A$ and $B$ such that
\begin{equation}
\label{sistemaab}
\left\{
\begin{array}{l}
\frac{A}{r_1^2}+\frac{B}{r_1^3}=\sum_{n=1}^{+\infty}\alpha_{n}r_1^{2n-1} \\[7pt]
\frac{A}{r_2^2}+\frac{B}{r_2^3}=\sum_{n=1}^{+\infty}\alpha_{n}r_2^{2n-1}
\end{array}
\right.
\end{equation}
By virtue of the low eccentricity assumption we have $r_1 \approx r_2 \approx r_0 \approx r$, where $r_0$ is the mean radius of an orbit. We can therefore solve system (\ref{sistemaab}) by considering the first order terms in (\ref{gint}):
\begin{equation}
\label{ab}
\left\{
\begin{array}{l}
A=4\alpha_1 r_0^3+6\alpha_2 r_0^5+8\alpha_3 r_0^7+...\\
B=-3\alpha_1 r_0^4-5\alpha_2 r_0^6-7\alpha_3 r_0^8-...\\
\end{array}
\right.
\end{equation}
From (\ref{ab}) we achieve $A$ and $B$ in powers of $r_0$:
\begin{equation}
\label{abserie}
\left\{
\begin{array}{l}
A=\sum_{n=1}^{+\infty}(2(n+1))\alpha_n r_0^{2n+1}=\sum_{n=1}^{+\infty}A_n \\[7pt]
B=-\bigl(\sum_{n=1}^{+\infty}(2n+1)\alpha_n r_0^{2(n+1)}\bigl)=-\sum_{n=1}^{+\infty}B_n
\end{array}
\right.
\end{equation}
Finally, using condition $r_1 \approx r_2 \approx r$, we can write $g_{int}(r)$ as
\begin{equation}
\label{gintsimpl}
g_{int}(r)=\frac{A}{r^2}+\frac{B}{r^3}.
\end{equation}
We see that this method produces a reduction of the series of radial functions (\ref{gint}) into the sum of two radial functions (\ref{gintsimpl}). Furthermore, the series of functions is simplified into the two numerical series (\ref{abserie}). From a 
physical point of view, the two contributions in the central field (\ref{gintsimpl}) have to be related respectively to the modification of the form of the Keplerian orbit and to the onset of the precession motion.
As we can see, the approximation (\ref{gintsimpl}) corresponds to the effect of rotation of the Keplerian orbit discussed in the previous Section.

\paragraph{External field}
In this case, by expanding the integrand function in (\ref{gravitpot}) in powers of $\frac{R}{r}$ and by remarking the invariance of the same formula with respect to the replacement $r \leftrightarrow R$, we obtain
$$
V(r)=-\frac{Gm}{r}\biggl(1+\sum_{n=1}^{+\infty}\biggl(\frac{(2n-1)!!}{(2n)!!}\biggl)^2 \biggl(\frac{R}{r}\biggl)^{2n}\biggl).
$$
Thus the gravitational field $g$ for $r>R$ is

\begin{eqnarray}
\nonumber
g_{ext}(r)&=&-\frac{dV(r)}{dr}
=-\frac{Gm}{R^2}\biggl(\biggl(\frac{R}{r}\biggl)^2\\
\nonumber
&+&\sum_{n=1}^{+\infty}\biggl(\frac{(2n-1)!!}{(2n)!!}\biggl)^2 (2n+1)\biggl(\frac{R}{r}\biggl)^{2(n+1)}\biggl)\\
\label{gest}
&=&\frac{\beta_1}{r^2}+\sum_{n=1}^{+\infty} \frac{\beta_{n+1}}{r^{2(n+1)}}.
\end{eqnarray}
Using the same method employed for the internal field, one can write (\ref{gest}) as
\begin{equation}
\label{gestsimpl}
g_{ext}(r)=\frac{C}{r^2}+\frac{D}{r^3}
\end{equation}
where the coefficients are
\begin{equation}
\label{cdserie}
\left\{
\begin{array}{l}
C=\beta_1-\sum_{n=1}^{+\infty}(2n-1) \frac{\beta_{n+1}}{r_0^{2n}}=C_0-\sum_{n=1}^{+\infty}C_n \\[7pt]
D=\sum_{n=1}^{+\infty}(2n) \frac{\beta_{n+1}}{r_0^{2n-1}}=-\sum_{n=1}^{+\infty}D_n.
\end{array}
\right.
\end{equation}
The result is formally the same as in the case of internal field and we have only to take into account apart the different values of the coefficients. 

\subsection{Calculation of precession frequency}

\noindent
We move forward now to the calculation of precession frequency, by using the techniques and the results performed so far.
The total force acting on a planet of mass $M$ is given by the sum of gravitational forces exerted by the Sun (with mass $M_{\odot}$ hereafter) and by the rings, representing the rest of planets. We have to consider that there are some rings inside planetary orbit and others outside, so that the total radial force is

\begin{eqnarray*}
F(r)&=&-\frac{GM_{\odot}M}{r^2}\\
&+&M\biggl(\sum_{j, internal} g_{int,j}(r)+\sum_{j, external} g_{ext,j}(r)\biggl)
\end{eqnarray*}
where the first sum is extended over the internal rings and the second over the external ones. Setting $k=GM_{\odot}M$ and using (\ref{abserie}),(\ref{gintsimpl}), (\ref{gestsimpl}) and (\ref{cdserie}) we obtain
\begin{eqnarray}
\nonumber
F(r)&=& \frac{-k+M(\sum_{j,internal}C_j+\sum_{j,external}A_j)}{r^2}\\
\label{fr}
&+&\frac{M(\sum_{j,internal}D_j+\sum_{j,external}B_j)}{r^3}\\
\nonumber
&=&\frac{-k+{\mit \Phi}}{r^2}+\frac{H}{r^3}.
\end{eqnarray}
As we can see, only the second term $\frac{H}{r^3}$ is responsible for precession, while the first one $\frac{\mit \Phi}{r^2}$ causes simply a change in eccentricity. Thus, the ring model is able to explain not only the apsidal precession but also the secular motion of eccentricity variation. By using (\ref{fr}) the orbit equation takes the form
\begin{equation}
\label{orbitering}
\frac{d^2 \rho}{d\theta^2} + \rho =\frac{kM}{||\textbf{L}||^2}-\frac{{\mit \Phi}M}{||\textbf{L}||^2} -\frac{HM\rho}{||\textbf{L}||^2}
\end{equation}
where $\rho=\frac{1}{r}$. In order to introduce a perturbation approach we define 
\begin{equation}
\label{eps}
\epsilon = \frac{\mit \Phi}{k}\backsim \frac{\sum_{i=1}^7 m_i}{M_{\odot}}\backsim 10^{-3} 
\end{equation}
where the mass $m_i$ is the mass of a ring (i.~e.~of a planet). In this way, equation (\ref{orbitering}) is written as
$$
\frac{d^2 \rho}{d\theta^2} + \rho =\frac{kM}{||\textbf{L}||^2}-\epsilon\frac{kM}{||\textbf{L}||^2}\biggl(1 +\frac{H}{\mit \Phi}\rho\biggl)
$$
which is in turn equivalent to the following system:
\begin{equation}
\label{pertsist}
\left\{
\begin{array}{l}
\frac{ds}{d\theta}=\frac{kM}{||\textbf{L}||^2}-\rho-\epsilon\frac{kM}{||\textbf{L}||^2}\biggl(1 +\frac{H}{\mit \Phi}\rho\biggl)\\
s=\frac{d\rho}{d\theta}
\end{array}
\right.
\end{equation}
which comes under the standard form in perturbation theory.
The solutions of the unperturbed problem $\epsilon=0$ are the ones of the Kepler problem:
\begin{equation*}
\left\{
\begin{array}{l}
\rho(\theta)=\frac{Mk}{||\textbf{L}||^2}(1+e\cos(\theta-\omega))=\frac{Mk}{||\textbf{L}||^2}(1+e\cos(f)) \\ \\
s(\theta)=-\frac{Mk}{||\textbf{L}||^2}e\sin(f)
\end{array}
\right.
\end{equation*}
(we set $f=\theta -\omega$).

\noindent
Applying the method of variation of constants, we replace the just written unperturbed solutions in (\ref{pertsist}). The integration constants $\omega$ and $e$ of the unperturbed problem now depend on the variable $\theta$. Straightforward calculations lead to the following differential system in $\omega$ and $e$: 
\begin{equation}
\label{pertsist2}
\left\{
\begin{array}{l}
\frac{de}{d\theta}\cos f +e\frac{d\omega}{d\theta}\sin f=0 \\[.5pt]
\frac{de}{d\theta}\sin f+e\bigl(1-\frac{d\omega}{d\theta}\bigl)\cos f\\[.5pt]
=e\cos f+\epsilon\bigl(1+\frac{H}{\mit \Phi}\frac{Mk}{||\textbf{L}||^2}(1+e\cos f)\bigl)
\end{array}
\right.
\end{equation} 
Combining the first equation in (\ref{pertsist2}) with the second one, we get
\begin{equation}
\label{comb}
\frac{d\omega}{d\theta}=-\frac{\epsilon}{e}\cos f \biggl(1+\frac{H}{\mit \Phi}\frac{Mk}{||\textbf{L}||^2}(1+e\cos f)\biggl).
\end{equation}
At this point we can employ the first order perturbation theory: by integrating (\ref{comb}) over an entire revolution period, we achieve
\begin{eqnarray*}
\Delta \omega&=&\omega(2\pi)-\omega(0)\\
&=&-\int_0^{2\pi}\frac{\epsilon}{e}\cos f \biggl(1+\frac{H}{\mit \Phi}\frac{Mk}{||\textbf{L}||^2}(1+e\cos f)\biggl)d\theta.
\end{eqnarray*}
Having in mind that $e$ and $\omega$ assume at the first order the constant unperturbed values, we have 
\begin{equation*}
\Delta \omega=-\frac{\epsilon \pi H}{\mit \Phi}\frac{Mk}{||\textbf{L}||^2}
\end{equation*}
and, recalling (\ref{eps}):
\begin{equation}
\label{delta}
\Delta \omega=-\frac{HM\pi}{||\textbf{L}||^2}=-\frac{HM}{4\pi M^2 r_0^4 }T_{riv}^2.
\end{equation}
The result depends on $H$ but not on ${\mit \Phi}$, consistently with the fact that the term bringing about precession is $\frac{H}{r^3}$. We also remark in (\ref{delta}) the inverse proportionality with respect to the square of the moment of inertia: the fact is physically reasonable, since the moment of inertia reveals the body's opposition to the motion of rotation, in this case the rotation of the apsidal line. Furthermore, the  proportionality with respect to $T_{riv}^2$ can be explained by considering that the greater is the revolution period, the longer is the time of exposure to the perturbation in a complete revolution orbit.
By means of (\ref{delta}) and assuming the slow advance of the apsidal line, we can write the precession frequency as
\begin{equation}
\label{pf}
\dot{\omega}_{th}=(\Delta \omega)|_{arcsec}\frac{1}{T_{riv}}\frac{3600}{2\pi}360(365\cdot 24\cdot 3600)\frac{s}{year}
\end{equation}
where the numerical factors let us convert the result from $\frac{rad}{s}$ to $\frac{arcsec}{year}$. 

\subsection{Planets precession frequencies}

\noindent
We now make use of (\ref{pf}) in order to calculate the precession frequencies of each planet. 
The calculation procedure takes into account both that the term $H$ in (\ref{delta}) varies from one planet to another and that it contains the specific values of each ring: actually, if we focus on the perturbation which  the planet with mass $M^{(i)}$ undergoes, the factor $H$ is
\begin{eqnarray}
\nonumber
& &H^{(i)}=-\\
& &M^{(i)}\biggl(\sum_{j,internal}
\sum_{n=1}^{+\infty}B_{j,n}^{(i)}+\sum_{j,external}\sum_{n=1}^{+\infty}D_{j,n}^{(i)} \biggl)
\label{hi}
\end{eqnarray}
where the $D_{j,n}^{(i)}$ and $B_{j,n}^{(i)}$ are respectively the contributions of the $j-th$ internal ring at the $n-order$ and of the $j-th$ external ring at the $n-order$. In order to calculate the precession frequencies using the expression (\ref{pf}) we cut the ring series at a certain order, depending on the accuracy which the precession frequencies are measured at. The precession frequencies data are acquired from \cite{PrecessionP} and plotted in Table 2. In the same text the frequencies are claimed to be with no error, thus we decide to take as errors on the frequencies, one of the last digits declared: for example the case of Mercury is treated as $(\dot{\bar{\omega}} \pm \Delta\dot{\bar{\omega}})_{obs}=5.75 \pm 0.01$. 

\end{multicols}

\begin{table}[!h]
\centering
\begin{tabular}{l*{8}{c}}
\toprule
& Mass & Mean radius($r_0$) & Revolution period($T_{riv}$)   \\
& (\textit{Kg}) & (\textit{m}) & (\textit{year}) \\
\midrule
Mercury & $3.302 \cdot 10^{23}$ & $5.79 \cdot 10^{10}$  & $0.241$  \\
Venus & $4.868 \cdot 10^{24}$ & $1.082\cdot 10^{11}$ & $0.615$ \\
Earth & $5.974 \cdot 10^{24}$ & $1.496\cdot 10^{11} $ & $1$ \\
Mars & $6.418 \cdot 10^{23}$ & $2.279\cdot 10^{11} $ & $1.88$ \\
Jupiter & $1.899 \cdot 10^{27}$ & $7.783\cdot 10^{11} $ & $11.86$ \\
Saturn & $5.685 \cdot 10^{26}$ & $1.429\cdot 10^{12} $ & $29.46$ \\
Uranus & $8.682 \cdot 10^{25}$ & $2.871\cdot 10^{12} $ & $84.10$ \\
Neptune & $1.024 \cdot 10^{26}$ & $4.498\cdot 10^{12} $ & $164.86$ \\
\bottomrule
\end{tabular}
\caption{Mass, mean radius and revolution period of the planets, according to the data stated in \cite{Universo}.}
\end{table}
\begin{table}[!h]
\centering
\begin{tabular}{l*{8}{c}}
\toprule
Terrestrial & $\dot{\bar{\omega}}_{obs}$ & $\bigl(\frac{\Delta\dot{\bar{\omega}}}{\dot{\bar{\omega}}}\bigl)_{obs}$  & Jovian &  $\dot{\bar{\omega}}_{obs}$ &$\bigl(\frac{\Delta\dot{\bar{\omega}}}{\dot{\bar{\omega}}}\bigl)_{obs}$ &\\
planets & ($\frac{arcsec}{year}$) & & planets &($\frac{arcsec}{year}$)& \\
\midrule
Mercury & 5.75 & 0.0017 & Jupiter & 6.55 & 0.0015 & \\ 
Venus & 2.05 & 0.005 & Saturn & 19.50 & 0.0005& \\ 
Earth & 11.45 & 0.0009 & Uranus & 3.44 & 0.003 &\\ 
Mars & 16.28 & 0.0006 & Neptune & 0.36 & 0.03 &\\ 
\bottomrule
\end{tabular}
\caption{Observed precession frequencies with relative precision (the data are taken from \cite{PrecessionP}).}
\end{table}

\begin{multicols}{2}

\noindent
Concerning the theoretical calculation of the frequencies and the cutting in the series (\ref{hi}), it is necessary to make the precision equal or as near as possible to the precision $\left(\frac{\Delta \dot{\bar{\omega}}}{\dot{\bar{\omega}}}\right)_{obs}$ of the measured frequencies.
Denoting by $\dot{\omega}_{1}$ an expressions of $\dot{\omega}_{th}$ which has $k$ terms and with  $\dot{\omega}_{2}$ and $\dot{\omega}_{3}$ two expressions of $\dot{\omega}_{th}$ which have respectively $k+1$ and $k+2$ terms, our criterion consists in checking if the following conditions are respected: 
$$
\begin{array}{l}
\bigl(\frac{\dot{\omega}_2-\dot{\omega}_1}{\dot{\omega}_1}\bigl)_{th}=\bigl(\frac{\Delta \dot{\omega}}{\dot{\omega}}\bigl)_{1,th}\gtrsim \bigl(\frac{\Delta \dot{\bar{\omega}}}{\dot{\bar{\omega}}}\bigl)_{oss} \\ 
\bigl(\frac{\dot{\omega}_3-\dot{\omega}_2}{\dot{\omega}_2}\bigl)_{th}=\bigl(\frac{\Delta \dot{\omega}}{\dot{\omega}}\bigl)_{2,th}< \bigl(\frac{\Delta \dot{\bar{\omega}}}{\dot{\bar{\omega}}}\bigl)_{oss}
\end{array}
$$
and in that case we set $\dot{\omega}_{th}=\dot{\omega}_{1,th}$. Calculations can be simplified if we recall (\ref{delta}) and (\ref{pf}) and we write
\begin{equation}
\label{approx}
\biggl(\frac{\dot{\omega}_2-\dot{\omega}_1}{\dot{\omega}_1}\biggl)_{th}=\frac{H_2-H_1}{H_1}
\end{equation}
where $H_1$ and $H_2$ are achieved from (\ref{hi}), by considering respectively $k$ and $k+1$ terms.
On the fround of (\ref{abserie}), (\ref{cdserie}) and (\ref{delta})--(\ref{hi}) we obtain the definitive  formula which we calculate the planets precession frequencies through
\begin{equation}
\label{formuladef}
\begin{array}{l}
\dot{\omega}^{(i)}_{th}=\frac{1}{4\pi {{r_0}^{(i)}}^4 }T_{riv}^{(i)}\biggl(\sum_{j,internal}\sum_{n=1}^{+\infty}B_{j,n}^{(i)}\\ \quad \quad +\sum_{j,external}\sum_{n=1}^{+\infty}D_{j,n}^{(i)} \biggl)N\\
B_{j,n}^{(i)}=Gm_j(2n+1)(2n)\bigl(\frac{(2n-1)!!}{(2n)!!}\bigl)^2\bigl(\frac{r_0^{(i)}}{R_j}\bigl)^{2n+1}r_0^{(i)} \\[7pt]
D_{j,n}^{(i)}=Gm_j(2n+1)(2n)\bigl(\frac{(2n-1)!!}{(2n)!!}\bigl)^2\bigl(\frac{R_j}{r_0^{(i)}}\bigl)^{2n-1}R_j.
\end{array}
\end{equation}
In (\ref{formuladef}) the coefficient $N$ is just needed for converting the result into $\frac{arcsec}{year}$ units. We point out the following quantities which play a significant role in calculating the precession frequency:
\begin{itemize}
\item Ring mass $m_j$: as we can see from the second and the third equation of (\ref{formuladef}), the precession frequency scales linearly with respect to the ring mass, so that at the lower orders the most massive rings give a contribution of the same importance as the closest ones.
\item The ratio between the mean radius of the planet $r_0$ and a ring radius $R_j$. The smaller is such ratio the closer are the orders of a ring series; moreover, several orders of this ring series appear in the precession frequency. 
\item Ring radius $r_0$ of the planet: the first equation in (\ref{formuladef}) shows that the precession frequency scales as $\frac{1}{{r_0^{(i)}}^4}$: the reason is related with our comment on the moment of inertia, made at the end of Section $\quad \quad$. So, much closer is the planet to the sun then much smaller is the opposition to the apsidal line precession.
\item Revolution period $T_{rev}$: again the first of (\ref{formuladef}) exhibits a linear dependence of the precession frequency on $T_{rev}$; an explication in this respect is provided at the end of the Section $\quad \quad$. 
\end{itemize}

\noindent
We are going to calculate the precession frequencies of each planet by employing the method illustrated so far and we will explain the difference among planets according to the scaling factors.

\noindent
The numerical values of the coefficients $B_{j,n}$ and $D_{j,n}$ for the planets are listed in the addendum. The data concerning the Solar System are achieved from Table 1.
 
\paragraph{Terrestrial planets}
Concerning Mercury, Venus, Hearth and Mars, it is necessary to hold many terms of (\ref{hi}) in order to get close the values of Table 2 it was necessary holding many terms in (\ref{hi}). 

\noindent
The following expressions, listed according to increasing distance of the planet from the Sun, indicate (\ref{hi}) for terrestrial planets, calculated via the approximation 
formulated in (\ref{approx}). 
\begin{eqnarray}
\nonumber
H^{(1)}=&-&M^{(1)}(B_{5,1}^{(1)}+B_{2,1}^{(1)}+B_{2,2}^{(1)}+B_{3,1}^{(1)}+B_{2,3}^{(1)}+B_{3,2}^{(1)}\\
\nonumber
&+&B_{2,4}^{(1)}+B_{6,1}^{(1)}
+ B_{2,5}^{(1)}+B_{3,3}^{(1)}+B_{4,1}^{(1)}+B_{5,2}^{(1)})\\
\label{merc}
&=&-5.17 \cdot 10^{48} \frac{kg \cdot m^4}{s^2}, \\
\nonumber
H^{(2)}=&-&M^{(2)}(B_{5,1}^{(2)}+B_{3,1}^{(2)}+B_{3,2}^{(2)}+B_{3,3}^{(2)}+B_{3,4}^{(2)}+B_{3,5}^{(2)}\\
\nonumber
&+&B_{3,6}^{(2)}+B_{3,7}^{(2)}+ B_{6,1}^{(2)}+B_{5,2}^{(2)}+B_{3,8}^{(2)}+B_{3,9}^{(2)})\\
\nonumber
&=&-7.71 \cdot 10^{50} \frac{kg \cdot m^4}{s^2}, \\
\nonumber
H^{(3)}=&-& M^{(3)}(B_{5,1}^{(3)}+D_{2,1}^{(3)}+D_{2,2}^{(3)}+D_{2,3}^{(3)}+D_{2,4}^{(3)}+B_{5,2}^{(3)}\\
\nonumber
&+&D_{2,5}^{(3)}+B_{6,1}^{(3)} +
D_{2,6}^{(3)}+D_{2,7}^{(3)}+D_{2,8}^{(3)}+B_{4,1}^{(3)}\\
\nonumber
&+&B_{4,2}^{(3)}+D_{2,9}^{(3)}+B_{4,3}^{(3)}
+ D_{2,10}^{(3)}+B_{4,4}^{(3)}+B_{5,3}^{(3)}\\
&+&D_{1,1}^{(3)}+D_{2,11}^{(3)})
\label{ter}=-2.33 \cdot 10^{51} \frac{kg \cdot m^4}{s^2}, \\
\nonumber
H^{(4)}=&-& M^{(4)}(B_{5,1}^{(4)}+B_{5,2}^{(4)}+D_{3,1}^{(4)}+B_{6,1}^{(4)}+D_{3,2}^{(4)}+D_{3,3}^{(4)}\\
&+&D_{2,1}^{(4)}
\nonumber
+ B_{5,3}^{(4)}+D_{3,4}^{(4)})\\
\label{mar}
&=&
-9.72 \cdot 10^{50} \frac{kg \cdot m^4}{s^2}.
\end{eqnarray}

\noindent
The terms are sorted by ascending order in magnitude.
An overview on the latter expression shows that the major contribution to the precession motion comes from the closest planets and from the most massive ones (as Jupiter and Saturn). In particular,  the first order of Jupiter's ring gives the largest contribution in all the computed expressions
$H^{(1)}$--$H^{(4)}$. In table 3 we list the precession frequencies calculated with the ring model ($\dot{\omega}_{teo}$) in comparison with the observed data.

\end{multicols}

\begin{table}[!h]
\centering
\begin{tabular}{l*{8}{c}}
\toprule
Planets & $\dot{\bar{\omega}}_{oss}$ & $\dot{\omega}_{teo}$  & $\bigl|\frac{\dot{\omega}_{teo}-\dot{\bar{\omega}}_{oss}}{\dot{\bar{\omega}}_{oss}}\bigl|$\\
& ($\frac{arcsec}{year}$)& ($\frac{arcsec}{year}$) & & \\
\midrule
Mercury & $5.75$ & $5.48$ & $0.05$  \\ 
Venus & $2.05$ & $11.61$ &   \\ 
Earth & $11.45$ & $12.68$ &  $0.11$\\ 
Mars & $16.28$ & $17.23$ & $0.06$ \\ 
\bottomrule
\end{tabular}
\caption{Theorical precession frequencies compared to observed data}
\end{table}

\begin{multicols}{2}

The calculated values are in agreement with the observed data -- eleven percent at worst -- except for Venus: as to the latter case, the model does not provide consistent results (actually the relative error has been omitted) owing to a low eccentricity of that planet (considerably lesser than the other ones) producing a more perturbability of the perihelion. perturbations. For such a case, the inadequacy of the calculated values in present also in \cite{PrecessionP}. Concerning the rest of the  planets, the differences among Mercury,Earth and Mars is strong and they can be explained on the ground of the scaling factors introduced in the previous Section:

\begin{itemize}
\item The ratio between the mean radii of Venus and Earth is $\frac{R_V}{R_T}=0.72$ and it is the largest ratio registered among internal or external planets. Immediately after the largest ratio is $\frac{R_T}{R_M}=0.66$ between Earth and Mars. The mass of Venus and of the Earth are an order of magnitude greater than those of Mars and Mercury. 
This is the reason why many perturbation terms of Venus appear in the expression (\ref{ter}) concerning the Earth, and many perturbation terms of the Earth appear in the expression (\ref{mar}) of Mars. Furthermore, all the terms which appearing (\ref{ter}) and in (\ref{mar}), are larger than the perturbation terms coming form Venus in the expression of Mercury (\ref{merc}).

\item The ratio between the mean radii of Earth and Jupiter $\frac{R_T}{R_G}=0.19$ and the ratio between the mean radii of Mars and Jupiter $\frac{R_M}{R_G}=0.29$ are larger than the ratio between the mean radius of Mercury and Jupiter $\frac{R_m}{R_G}=0.07$: actually $\frac{R_T}{R_G}=2.7 \frac{R_m}{R_G}$ and $\frac{R_M}{R_G}=4.1 \frac{R_m}{R_G}$. Therefore, Jupiter contributes to perturbing Earth and Mars to a greater extent than Mercury.
\item The external field due to the rings is stronger than the internal field, for the same radial distance from a point of the ring.
\item The revolution periods of Earth and Mars are greater than the revolution period of Mercury (third Kepler's law).
\end{itemize}
All these factors help us explain the disparities in the precession frequencies of the Terrestrial planets.
\paragraph{Jovian planets}
In order to achieve the degree of accuracy listed in Table 2 it is necessary to hold many terms in (\ref{hi}). The following expressions show (\ref{hi}) for each Jovian planet, and the stopping criterion is the one we explained above. As before, the values of $(H^{(i)}$ are listed by increasing distance from the Sun:
\begin{eqnarray}
\nonumber
H^{(5)}=&-&M^{(5)}(B_{6,1}^{(5)}+B_{6,2}^{(5)}+B_{6,3}^{(5)}+B_{6,4}^{(5)}+B_{6,5}^{(5)}+B_{7,1}^{(5)}\\
\nonumber
&+&B_{6,6}^{(5)}+B_{8,1}^{(5)})=-2.66 \cdot 10^{55} \frac{kg \cdot m^4}{s^2}, \\
\nonumber
H^{(6)}=&-&M^{(6)}(D_{5,1}^{(6)}+D_{5,2}^{(6)}+D_{5,3}^{(6)}+D_{5,4}^{(6)}+D_{5,5}^{(6)}+B_{7,1}^{(6)}\\
\nonumber
&+&D_{5,6}^{(6)}
+B_{7,2}^{(6)}+B_{8,1}^{(6)}
+D_{5,8}^{(6)}+B_{7,3}^{(6)}+D_{5,9}^{(6)})\\
\nonumber
&=&-9.06 \cdot 10^{55} \frac{kg \cdot m^4}{s^2},\\
\nonumber
H^{(7)}=&-&M^{(7)}(D_{6,1}^{(7)}+D_{5,1}^{(7)}+D_{6,2}^{(7)}+B_{8,1}^{(7)}+D_{6,3}^{(7)}\\
\nonumber
&+&B_{8,2}^{(7)}+D_{5,2}^{(7)}
+B_{8,3}^{(7)}+D_{6,4}^{(7)}
+B_{8,4}^{(7)}+B_{8,5}^{(7)}\\
\nonumber
&+&D_{6,5}^{(7)})
=-1.17 \cdot 10^{55} \frac{kg \cdot m^4}{s^2},\\
\nonumber
H^{(8)}=&-&M^{(8)}(D_{6,1}^{(8)}+D_{5,1}^{(8)}+D_{7,1}^{(8)}+D_{7,2}^{(8)}+D_{7,3}^{(8)}\\
\nonumber
&+&D_{6,2}^{(8)})=
-9.38 \cdot 10^{54} \frac{kg \cdot m^4}{s^2}.
\end{eqnarray}
Again, 
the terms in the sums are sorted by ascending order of magnitude.
Concerning the Jovian planets the contribute to the precession mainly comes from the mutual interaction among themselves, since they are much more massive than the terrestrial planets. As a consequence, the values of $H^{(i)}$ are much larger than those computed for the terrestrial set. 
The precession frequencies for the Jovian planets calculated via the ring model are listed in Table 4 
and compared to the observed data.

\begin{table}[!h]
\centering
\begin{tabular}{l*{8}{c}}
\toprule
Planets & $\dot{\bar{\omega}}_{oss}$ & $\dot{\omega}_{teo}$  & $\bigl|\frac{\dot{\omega}_{teo}-\dot{\bar{\omega}}_{oss}}{\dot{\bar{\omega}}_{oss}}\bigl|$\\
& ($\frac{arcsec}{year}$)& ($\frac{arcsec}{year}$) & & \\
\midrule
Jupiter & $6.55$ & $7.40$ & $0.13$  \\ 
Saturn & $19.50$ & $18.39$ & $0.06$  \\ 
Uranus & $3.34$ & $2.71$ &  $0.19$\\ 
Neptune & $0.36$ & $0.60$ & $0.7$ \\ 
\bottomrule
\end{tabular}
\caption{Theorical precession frequencies compared to observed data}
\end{table}

\noindent
As one can remark, the attained results fit with the observed data --at worst twenty percent of precision -- with the exception of Neptune, which exhibits a precision of seventy percent. Furthermore, the differences among the precession frequencies are remarkable. Such differences can be explained by considering once again the scaling factors:
\begin{itemize}
\item The largest ratio that the mean radius of Jupiter forms with any other planet is the one with Saturn $\frac{R_J}{R_S}$ is the greatest ratio between internal and external planets including Jupiter; on the other hand, the mass of Saturn is the second one in the entire solar system. These two facts explain 
the maximum factor $(\ref{hi})$ for Saturn in the overall Solar system.

\item The huge mean radii of the two external planets Uranus and Neptune
produce the lowest precession frequencies. 
\end{itemize}

\section{Corrections to the ring model}

We will give here an explication of the disparity between the theorical results obtained via the ring model and the observed data. As we already stated, the model is not suitable for the case of Venus because of its eccentricity: we may improve the model by introducing a non zero eccentricity for the orbit.
Secondly, the assumption of complanar orbits can be released too, in order to make way for a model providing more realistic data.
In line with this thinking, one can set the problem of motion by employing the classical mechanics which  encompasses eccentricity and relative inclination of the orbit planes and at the same time by still making use of the rings assumption to formulate perturbation. The geometrical features of the new problem are more complex and a system of Keplerian coordinates, whose details can be found in \cite{Precession2}, are needed.

\noindent
It can be seen that the improvement produced by the correction leads to a better accuracy and to a concrete  closeness of the calculated precession frequencies to the observed data, except for Mercury. In this latter case, the theorical value is lower than the one calculated by the not corrected model and the difference from the observed data is more remarkable: more precisely, the theorical precession frequency of Mercury becomes $5.32\frac{arcsec}{year}$ versus the observed datum $5.75\frac{arcsec}{year}$ (\cite{Astronomia}). The disparity can be explained by the proximity to the Sun and the non negligible curvature of the space--time, so that the relavistic effect gives rise to an additional term in the total force:
\begin{equation*}
\textbf{F}_{tot}=-\frac{k}{r^2}\frac{\textbf{r}}{||\textbf{r}||}+\textbf{F}_{rings}-\frac{3GM_{\odot}}{c^2 r^4}\frac{\textbf{r}}{||\textbf{r}||}
\end{equation*}
where the first term is the contribution of the central Keplerian force and the second term takes account of the effects due to the planets. The latter one is still a classical term but no longer of central type, owing to the more complex geometrical structure of the problem. The last term is the relativistic contribution of space--time curvature caused by the Sun mass. Hence, the total perturbation to the equation of motion is
\begin{equation}
\label{nuovapert}
\textbf{F}=\textbf{F}_{rings}-\frac{3GM_{\odot}}{c^2 r^4}\frac{\textbf{r}}{||\textbf{r}||}
\end{equation}  
As shown above, the precession frequency induced merely by the first and the second term in (\ref{nuovapert}) is given by the sum of the two frequencies corresponding to each of the contributions.
Once calculations have been carried out, it can be seen that the relativistic term is able to explain the difference between the theorical value and the observed one $0.43 \frac{arcsec}{year}$.

\section{Conclusions}
The Lenz vector shows that the ordinary reason for the precession motion is the perturbation to the keplerian motion produced by planets. Actually, the Lenz vector is a constant of motion for the Keplerian motion. For a single planet, the other ones have the effect of an additional force acting on it and this entails that Lenz vector is no longer a constant of motion but it precedes around a given axis according to a specific frequency. On that basis, we formulated the ring model in order to calculate the planets precession frequencies by means of a perturbation method.

The model consists of a central force plus a term $\frac{H}{r^3}$ which causes the rotation of the apsidal lines in the revolution orbit plane.
The ring model provide consistent results for any planet except for Venus: actually Venus is much more susceptible to perturbations because of its eccentricity, so that the method is not quite reliable for describing the precession motion. 
The model shows which are the prevalent factors in the precession frequency: the ring mass $m_j$, the ratio between the mean radius of the planet $r_0$ and the ring radius $R_j$, the ring radius $r_0$, the revolution period $T_{rev}$. By evaluating such factors for each planet we can explain the dissimilarities in precession frequencies attained by the ring model.

We stressed the disadvantage in the central force approximation 
of not correctly performing the relativistic effect on the apsidal precession of Mercury: as a matter of facts, the relative error between the theorical value and the observed data exhibits the same order for Mercury and for the rest of the planets, with the exception of Venus. Such a limit clearly appears if one releases the assumptions of complanar and circular orbits. The relativistic correction can be summed up to the classic term in order to obtain the correct precession frequency for Mercury.

The observed values fit adequately with the theorical values obtained with the ring model. Furthermore, leaving the basic assumptions of the ring model produces susceptibility to the relativistic effect only for one planet, with a correction of eight percentç we can conclude that the major contribution to the precession motion has to be ascribed to the classical effect.

\section{Addendum: Coefficients numerical values $B_{j,n}$ and $D_{j,n}$}

\noindent
We list below the numerical values of the coefficients $B_{j,n}$ e $D_{j,n}$, calculated using Matlab, for each planet of the Solar system. The units of the coefficients in the SI system arem 
$[B_{j,n}]=[D_{j,n}]=\frac{m^4}{s^2}$. 

\end{multicols}

{\footnotesize 
\paragraph{Mercury}
\begin{center}
\begin{tabular}{l*{8}{c}}
\toprule & $n=1$ & $n=2$ &  $n=3$ & $n=4$& $n=5$ & $n=6$ & \\
$B_{j,n}$ \\
\midrule
$B_{2,n}^{(1)}$ & $4.32 \cdot 10^{24}$ & $2.32 \cdot 10^{24}$ & $9.69 \cdot 10^{23}$ & $3.64 \cdot 10^{23}$ & $1.29 \cdot 10^{23}$ & $4.40 \cdot 10^{22}$ \\ 
$B_{3,n}^{(1)}$ & $2.01 \cdot 10^{24}$ & $5.64 \cdot 10^{23}$ & $1.23 \cdot 10^{23}$ & $2.42 \cdot 10^{22}$ & $4.49 \cdot 10^{21}$ & $8.01 \cdot 10^{20}$ \\ 
$B_{4,n}^{(1)}$ & $6.10 \cdot 10^{22}$ & $7.38 \cdot 10^{21}$ & $6.95 \cdot 10^{20}$ & $5.88 \cdot 10^{19}$ & $4.70 \cdot 10^{18}$ & $3.61 \cdot 10^{17}$ \\ 
$B_{5,n}^{(1)}$ & $4.53 \cdot 10^{24}$ & $4.70 \cdot 10^{22}$ & $3.79 \cdot 10^{20}$ & $2.76 \cdot 10^{18}$ & $1.89 \cdot 10^{16}$ & $1.24 \cdot 10^{14}$ \\
$B_{6,n}^{(1)}$ & $2.19 \cdot 10^{23}$ & $6.74 \cdot 10^{20}$ & $1.61 \cdot 10^{18}$ & $3.48 \cdot 10^{15}$ & $7.07 \cdot 10^{12}$ & $1.38 \cdot 10^{10}$ \\
$B_{7,n}^{(1)}$ & $4.13 \cdot 10^{21}$ & $3.15 \cdot 10^{18}$ & $1.87 \cdot 10^{15}$ & $9.96 \cdot 10^{11}$ & $5.01 \cdot 10^{8}$ & $2.43 \cdot 10^{5}$ \\
$B_{8,n}^{(1)}$ & $1.27 \cdot 10^{21}$ & $3.93 \cdot 10^{17}$ & $9.50 \cdot 10^{13}$ & $2.07 \cdot 10^{10}$ & $4.24 \cdot 10^{6}$ & $8.36 \cdot 10^{2}$ \\ 
\bottomrule
\end{tabular}
\end{center}

\paragraph{Venus}
\begin{center}
\begin{tabular}{l*{8}{c}}
\toprule
 & $n=1$ & $n=2$ & $n=3$ & $n=4$ & $n=5$ & \\
\midrule
$B_{3,n}^{(2)} $ & $2.45 \cdot 10^{25}$ & $2.40 \cdot 10^{25}$ & $1.83 \cdot 10^{25}$ & $1.26 \cdot 10^{25}$ & $8.14 \cdot 10^{24}$ & \\
$B_{4,n}^{(2)}$ & $7.44 \cdot 10^{23}$ & $3.14 \cdot 10^{23}$ & $1.03 \cdot 10^{23}$ & $3.06 \cdot 10^{22}$ & $8.52 \cdot 10^{21}$ &\\
$B_{5,n}^{(2)}$ & $5.52 \cdot 10^{25}$ & $2.00 \cdot 10^{24}$ & $5.64 \cdot 10^{22}$ & $1.43 \cdot 10^{21}$ & $3.42 \cdot 10^{19}$ & \\
$B_{6,n}^{(2)}$ & $2.67 \cdot 10^{24}$ & $2.87 \cdot 10^{22}$ & $2.40 \cdot 10^{20}$ & $1.81 \cdot 10^{18}$ & $1.28 \cdot 10^{16}$ & \\
$B_{7,n}^{(2)}$ & $5.03 \cdot 10^{22}$ & $1.34 \cdot 10^{20}$ & $2.78 \cdot 10^{17}$ & $5.17 \cdot 10^{14}$ & $9.09 \cdot 10^{11}$ & \\
$B_{8,n}^{(2)}$ & $1.54 \cdot 10^{22}$ & $1.67 \cdot 10^{19}$ & $1.41 \cdot 10^{16}$ & $1.07 \cdot 10^{13}$ & $7.68 \cdot 10^{9}$ & \\
\bottomrule
\end{tabular}
\end{center}

\begin{center}
\begin{tabular}{l*{8}{c}}
\toprule
 & $n=6$ & $n=7$ & $n=8$ & $n=9$ & \\
\midrule
$B_{3,n}^{(2)} $ & $5.07 \cdot 10^{24}$ & $3.08 \cdot 10^{24}$ & $1.83 \cdot 10^{24}$ & $1.08 \cdot 10^{24}$ & \\
$B_{4,n}^{(2)}$ & $2.29 \cdot 10^{21}$ & $5.99 \cdot 10^{20}$ & $1.54 \cdot 10^{20}$ & $3.89 \cdot 10^{19}$ & \\
$B_{5,n}^{(2)}$ & $7.88 \cdot 10^{17}$ & $1.77 \cdot 10^{16}$ & $3.89 \cdot 10^{14}$ & $8.43 \cdot 10^{12}$ & \\
$B_{6,n}^{(2)}$ & $8.76 \cdot 10^{13}$ & $5.83 \cdot 10^{11}$ & $3.80 \cdot 10^{9}$ & $2.45 \cdot 10^{7}$ & \\
$B_{7,n}^{(2)}$ & $1.54 \cdot 10^{9}$ & $2.54 \cdot 10^{6}$ & $4.10 \cdot 10^{3}$ & $6.53$ & \\
$B_{8,n}^{(2)}$ & $5.30 \cdot 10^{6}$ & $3.56 \cdot 10^{3}$ & $2.34$ & $1.52 \cdot 10^{-3}$ & \\
\bottomrule
\end{tabular}
\end{center}

\paragraph{Earth}
\begin{center}
\begin{tabular}{l*{8}{c}}
\toprule
 & $n=1$ & $n=2$ & $n=3$ & $n=4$ & $n=5$ & $n=6$ & \\
\midrule
$D_{1,n}^{(3)}$ & $7.40 \cdot 10^{23}$ & $2.08 \cdot 10^{23}$ & $4.54 \cdot 10^{22}$ & $8.93 \cdot 10^{21}$ & $1.66 \cdot 10^{21}$ & $2.95 \cdot 10^{20}$  \\
$D_{2,n}^{(3)}$ & $3.81 \cdot 10^{25}$ & $3.74 \cdot 10^{25}$ & $2.85 \cdot 10^{25}$ & $1.96 \cdot 10^{25}$ & $1.27 \cdot 10^{25}$ & $7.90 \cdot 10^{24}$  \\
\toprule
& $n=7$ & $n=8$ & $n=9$ & $n=10$ & $n=11$ & $n=12$ \\
\midrule
$D_{1,n}^{(3)}$ & $5.14 \cdot 10^{19}$ & $8.76 \cdot 10^{18}$ & $1.47 \cdot 10^{18}$ & $2.44 \cdot 10^{17}$ & $4.02 \cdot 10^{16}$ & $6.55 \cdot 10^{15}$  \\
$D_{2,n}^{(3)}$ & $4.80 \cdot 10^{24}$ & $2.86 \cdot 10^{24}$ & $1.68 \cdot 10^{24}$ & $9.72 \cdot 10^{23}$ & $5.58 \cdot 10^{23}$ & $3.18 \cdot 10^{23}$\\
\bottomrule
\end{tabular}
\end{center}

\begin{center}
\begin{tabular}{l*{8}{c}}
\toprule
 & $n=1$ & $n=2$ & $n=3$ & $n=4$ & $n=5$ & $n=6$ & \\
\midrule
$B_{4,n}^{(3)}$ & $2.72 \cdot 10^{24}$ & $2.20 \cdot 10^{24}$ & $1.38 \cdot 10^{24}$ & $7.80 \cdot 10^{23}$ & $4.16 \cdot 10^{23}$ & $2.14 \cdot 10^{23}$ \\
$B_{5,n}^{(3)}$ & $2.02 \cdot 10^{26}$ & $1.40 \cdot 10^{25}$ & $7.53 \cdot 10^{23}$ & $3.65 \cdot 10^{22}$ & $1.67 \cdot 10^{21}$ & $7.35 \cdot 10^{19}$ \\
$B_{6,n}^{(3)}$ & $9.76 \cdot 10^{24}$ & $2.01 \cdot 10^{23}$ & $3.21 \cdot 10^{21}$ & $4.61 \cdot 10^{19}$ & $6.26 \cdot 10^{17}$ & $8.17 \cdot 10^{15}$ \\
$B_{7,n^{(3)}}$ & $1.84 \cdot 10^{23}$ & $9.36 \cdot 10^{20}$ & $3.71 \cdot 10^{18}$ & $1.32 \cdot 10^{16}$ & $4.44 \cdot 10^{14}$ & $1.44 \cdot 10^{11}$ \\
$B_{8,n}^{(3)}$ & $5.64 \cdot 10^{22}$ & $1.17 \cdot 10^{20}$ & $1.89 \cdot 10^{17}$ & $2.74 \cdot 10^{14}$ & $3.75 \cdot 10^{21}$ & $4.94 \cdot 10^{8}$ \\
\bottomrule
\end{tabular}
\end{center}
\paragraph{Mars}
\begin{center}
\begin{tabular}{l*{8}{c}}
\toprule
 & $n=1$ & $n=2$ & $n=3$ & $n=4$ & $n=5$ & $n=6$ & \\
\midrule
$B_{5,n}^{(4)}$ & $1.09 \cdot 10^{27}$ & $1.75 \cdot 10^{26}$ & $2.19 \cdot 10^{25}$ & $2.46 \cdot 10^{24}$ & $2.61 \cdot 10^{23}$ & $2.67 \cdot 10^{22}$ \\
$B_{6,n}^{(4)}$ & $5.26 \cdot 10^{25}$ & $2.51 \cdot 10^{24}$ & $9.30 \cdot 10^{22}$ & $3.11 \cdot 10^{21}$ & $9.77 \cdot 10^{19}$ & $2.96 \cdot 10^{18}$ \\
$B_{7,n}^{(4)}$ & $9.90 \cdot 10^{23}$ & $1.17 \cdot 10^{22}$ & $1.08 \cdot 10^{20}$ & $8.89 \cdot 10^{17}$ & $6.93 \cdot 10^{15}$ & $5.21 \cdot 10^{13}$ \\
$B_{8,n}^{(4)}$ & $3.04 \cdot 10^{23}$ & $1.46 \cdot 10^{21}$ & $5.47 \cdot 10^{18}$ & $1.84 \cdot 10^{16}$ & $5.86 \cdot 10^{13}$ & $1.79 \cdot 10^{11}$ \\
\bottomrule
\end{tabular}
\end{center}

\begin{center}
\begin{tabular}{l*{8}{c}}
\toprule
 & $n=1$ & $n=2$ & $n=3$ & $n=4$ & $n=5$ & $n=6$ & \\
\midrule
$D_{1,n}^{(4)}$ & $4.86 \cdot 10^{23}$ & $5.88 \cdot 10^{22}$ & $5.54 \cdot 10^{21}$ & $4.69 \cdot 10^{20}$ & $3.75 \cdot 10^{19}$ & $2.88 \cdot 10^{18}$  \\
$D_{2,n}^{(4)}$ & $2.50 \cdot 10^{25}$ & $1.06 \cdot 10^{25}$ & $3.48 \cdot 10^{24}$ & $1.03 \cdot 10^{24}$ & $2.87 \cdot 10^{23}$ & $7.70 \cdot 10^{22}$  \\
$D_{3,n}^{(4)}$ & $5.87 \cdot 10^{25}$ & $4.74 \cdot 10^{25}$ & $2.98 \cdot 10^{25}$ & $1.69 \cdot 10^{25}$ & $8.99 \cdot 10^{24}$ & $4.61 \cdot 10^{24}$  \\
\toprule
& $n=7$ & $n=8$ & $n=9$ & $n=10$ & $n=11$ & $n=12$ \\
\midrule
$D_{1,n}^{(4)}$ & $2.16 \cdot 10^{17}$ & $1.59 \cdot 10^{16}$ & $1.15 \cdot 10^{15}$ & $8.21 \cdot 10^{13}$ & $5.82 \cdot 10^{12}$ & $4.09 \cdot 10^{11}$  \\
$D_{2,n}^{(4)}$ & $2.02 \cdot 10^{22}$ & $5.17 \cdot 10^{21}$ & $1.31 \cdot 10^{21}$ & $3.27 \cdot 10^{20}$ & $8.08 \cdot 10^{19}$ & $1.98 \cdot 10^{19}$\\
$D_{3,n}^{(4)}$ & $2.31 \cdot 10^{24}$ & $1.13 \cdot 10^{24}$ & $5.47 \cdot 10^{23}$ & $2.61 \cdot 10^{23}$ & $1.24 \cdot 10^{23}$ & $5.80 \cdot 10^{22}$  \\
\bottomrule
\end{tabular}
\end{center}

\paragraph{Jupiter}
\begin{center}
\begin{tabular}{l*{8}{c}}
\toprule
 & $n=1$ & $n=2$ & $n=3$ & $n=4$ & $n=5$ \\
\midrule
$B_{6,n}^{(5)}$ & $7.15 \cdot 10^{27}$ & $3.98 \cdot 10^{27}$ & $1.72 \cdot 10^{27}$ & $6.70 \cdot 10^{26}$ & $2.46 \cdot 10^{26}$ \\
$B_{7,n}^{(5)}$ & $1.35 \cdot 10^{26}$ & $1.86 \cdot 10^{25}$ & $1.99 \cdot 10^{24}$ & $1.92 \cdot 10^{23}$ & $1.74 \cdot 10^{22}$ \\
$B_{8,n}^{(5)}$ & $4.13 \cdot 10^{25}$ & $2.32 \cdot 10^{24}$ & $1.01 \cdot 10^{23}$ & $3.98 \cdot 10^{21}$ & $1.47 \cdot 10^{20}$ \\
\toprule
& $n=6$ & $n=7$ & $n=8$ & $n=9$ & $n=10$ \\
\midrule
$B_{6,n}^{(5)}$ & $8.69 \cdot 10^{25}$ & $2.99 \cdot 10^{25}$ & $1.01 \cdot 10^{25}$ & $3.36 \cdot 10^{24}$ & $1.11 \cdot 10^{24}$ \\
$B_{7,n}^{(5)}$ & $1.53 \cdot 10^{21}$ & $1.30 \cdot 10^{20}$ & $1.09 \cdot 10^{19}$ & $8.99 \cdot 10^{17}$ & $7.32 \cdot 10^{16}$ \\
$B_{8,n}^{(5)}$ & $5.26 \cdot 10^{18}$ & $1.83 \cdot 10^{17}$ & $6.23 \cdot 10^{15}$ & $2.09 \cdot 10^{14}$ & $6.94 \cdot 10^{12}$ \\
\bottomrule
\end{tabular}
\end{center}

\begin{center}
\begin{tabular}{l*{8}{c}}
\toprule
 & $n=1$ & $n=2$ & $n=3$ \\
\midrule
$D_{1,n}^{(5)}$ & $1.42 \cdot 10^{23}$ & $1.48 \cdot 10^{21}$ & $1.19 \cdot 10^{19}$ \\
$D_{2,n}^{(5)}$ & $7.33 \cdot 10^{24}$ & $2.65 \cdot 10^{23}$ & $7.48 \cdot 10^{21}$ \\
$D_{3,n}^{(5)}$ & $1.72 \cdot 10^{25}$ & $1.19 \cdot 10^{24}$ & $6.42 \cdot 10^{22}$ \\
$D_{4,n}^{(5)}$ & $4.29 \cdot 10^{24}$ & $6.89 \cdot 10^{23}$ & $8.61 \cdot 10^{22}$ \\
\bottomrule
\end{tabular}
\end{center}

\paragraph{Saturn}

\begin{center}
\begin{tabular}{l*{8}{c}}
\toprule
 & $n=1$ & $n=2$ & $n=3$ & $n=4$ & $n=5$ & $n=6$ & \\
\midrule
$B_{7,n}^{(6)}$ & $1.53 \cdot 10^{27}$ & $7.11 \cdot 10^{26}$ & $2.57 \cdot 10^{26}$ & $8.35 \cdot 10^{25}$ & $2.56 \cdot 10^{25}$ & $7.56 \cdot 10^{24}$  \\
$B_{8,n}^{(6)}$ & $4.69 \cdot 10^{26}$ & $8.88 \cdot 10^{25}$ & $1.31 \cdot 10^{25}$ & $1.73 \cdot 10^{24}$ & $2.16 \cdot 10^{23}$ & $2.60 \cdot 10^{22}$  \\
\bottomrule
\end{tabular}
\end{center}

\begin{center}
\begin{tabular}{l*{8}{c}}
\toprule
 & $n=1$ & $n=2$ & $n=3$ & $n=4$ & $n=5$ \\
\midrule
$D_{1,n}^{(6)}$ & $7.75 \cdot 10^{22}$ & $2.39 \cdot 10^{20}$ & $5.71 \cdot 10^{17}$ & $1.23 \cdot 10^{15}$ & $2.50 \cdot 10^{12}$ \\
$D_{2,n}^{(6)}$ & $3.99 \cdot 10^{24}$ & $4.29 \cdot 10^{22}$ & $3.59 \cdot 10^{20}$ & $2.70 \cdot 10^{18}$ & $1.91 \cdot 10^{16}$ \\
$D_{3,n}^{(6)}$ & $9.36 \cdot 10^{24}$ & $1.92 \cdot 10^{23}$ & $3.07 \cdot 10^{21}$ & $4.42 \cdot 10^{19}$ & $6.00 \cdot 10^{17}$ \\
$D_{4,n}^{(6)}$ & $2.33 \cdot 10^{24}$ & $1.11 \cdot 10^{23}$ & $4.13 \cdot 10^{21}$ & $1.38 \cdot 10^{20}$ & $4.34 \cdot 10^{18}$ \\
$D_{5,n}^{(6)}$ & $8.05 \cdot 10^{28}$ & $4.48 \cdot 10^{28}$ & $1.94 \cdot 10^{28}$ & $7.54 \cdot 10^{27}$ & $2.77 \cdot 10^{27}$ \\
\toprule
& $n=6$ & $n=7$ & $n=8$ & $n=9$ & $n=10$ \\
\midrule
$D_{1,n}^{(6)}$ & $4.89 \cdot 10^{9}$ & $9.32 \cdot 10^{6}$ & $1.74 \cdot 10^{4}$ & $3.21 \cdot 10$ & $5.84 \cdot 10^{-2}$ \\
$D_{2,n}^{(6)}$ & $1.31 \cdot 10^{14}$ & $8.70 \cdot 10^{11}$ & $5.68 \cdot 10^{9}$ & $3.65 \cdot 10^{7}$ & $2.32 \cdot 10^{5}$ \\
$D_{3,n}^{(6)}$ & $7.83 \cdot 10^{15}$ & $9.97 \cdot 10^{13}$ & $1.24 \cdot 10^{12}$ & $1.53 \cdot 10^{10}$ & $1.86 \cdot 10^{8}$ \\
$D_{4,n}^{(6)}$ & $1.31 \cdot 10^{17}$ & $3.88 \cdot 10^{15}$ & $1.12 \cdot 10^{14}$ & $3.21 \cdot 10^{12}$ & $9.04 \cdot 10^{10}$ \\
$D_{5,n}^{(6)}$ & $9.79 \cdot 10^{26}$ & $3.37 \cdot 10^{26}$ & $1.14 \cdot 10^{26}$ & $3.79 \cdot 10^{25}$ & $1.25 \cdot 10^{25}$ \\
\bottomrule
\end{tabular}
\end{center}

\paragraph{Uranus}

\begin{center}
\begin{tabular}{l*{8}{c}}
\toprule
 & $n=1$ & $n=2$ & $n=3$ & $n=4$ & $n=5$ \\
\midrule
$B_{8,n}^{(7)}$ & $7.65 \cdot 10^{27}$ & $5.84 \cdot 10^{27}$ & $3.47 \cdot 10^{27}$ & $1.86 \cdot 10^{27}$ & $9.36 \cdot 10^{26}$ \\
\toprule
& $n=6$ & $n=7$ & $n=8$ & $n=9$ & $n=10$ \\
\midrule
$B_{8,n}^{(7)}$ & $4.54 \cdot 10^{26}$ & $2.15 \cdot 10^{26}$ & $9.96 \cdot 10^{25}$ & $4.55 \cdot 10^{25}$ & $2.06 \cdot 10^{25}$ \\
\bottomrule
\end{tabular}
\end{center}
 \begin{center}
\begin{tabular}{l*{8}{c}}
\toprule
 & $n=1$ & $n=2$ & $n=3$ & $n=4$ & $n=5$ & $n=6$ \\
\midrule
$D_{1,n}^{(7)}$ & $3.86 \cdot 10^{22}$ & $2.94 \cdot 10^{19}$ & $1.74 \cdot 10^{16}$ & $9.31 \cdot 10^{12}$ & $4.69 \cdot 10^{9}$ & $2.27 \cdot 10^{6}$ \\
$D_{2,n}^{(7)}$ & $1.99 \cdot 10^{24}$ & $5.29 \cdot 10^{21}$ & $1.10 \cdot 10^{19}$ & $2.04 \cdot 10^{16}$ & $3.59 \cdot 10^{13}$ & $6.08 \cdot 10^{10}$ \\
$D_{3,n}^{(7)}$ & $4.66 \cdot 10^{24}$ & $2.37 \cdot 10^{22}$ & $9.39 \cdot 10^{19}$ & $3.35 \cdot 10^{17}$ & $1.12 \cdot 10^{15}$ & $3.64 \cdot 10^{12}$ \\
$D_{4,n}^{(7)}$ & $1.16 \cdot 10^{24}$ & $1.37 \cdot 10^{22}$ & $1.26 \cdot 10^{20}$ & $1.04 \cdot 10^{18}$ & $8.13 \cdot 10^{15}$ & $6.11 \cdot 10^{13}$ \\
$D_{5,n}^{(7)}$ & $4.01 \cdot 10^{28}$ & $5.52 \cdot 10^{27}$ & $5.92 \cdot 10^{26}$ & $5.71 \cdot 10^{25}$ & $5.19 \cdot 10^{24}$ & $4.55 \cdot 10^{23}$ \\
$D_{6,n}^{(7)}$ & $4.05 \cdot 10^{28}$ & $1.88 \cdot 10^{28}$ & $6.79 \cdot 10^{27}$ & $2.21 \cdot 10^{27}$ & $6.77 \cdot 10^{26}$ & $2.00 \cdot 10^{26}$ \\
\bottomrule
\end{tabular}
\end{center}

\paragraph{Neptune}
\begin{center}
\begin{tabular}{l*{8}{c}}
\toprule
 & $n=1$ & $n=2$ & $n=3$ & $n=4$ & $n=5$ & \\
\midrule
$D_{1,n}^{(8)} $ & $2.46 \cdot 10^{22}$ & $7.65 \cdot 10^{18}$ & $1.85 \cdot 10^{15}$ & $4.02 \cdot 10^{11}$ & $8.24 \cdot 10^{7}$ & \\
$D_{2,n}^{(8)}$ & $1.27 \cdot 10^{24}$ & $1.38 \cdot 10^{21}$ & $1.16 \cdot 10^{18}$ & $8.81 \cdot 10^{14}$ & $6.31 \cdot 10^{11}$ &\\
$D_{3,n}^{(8)}$ & $2.97 \cdot 10^{24}$ & $6.17 \cdot 10^{21}$ & $9.95 \cdot 10^{18}$ & $1.44 \cdot 10^{16}$ & $1.98 \cdot 10^{13}$ & \\
$D_{4,n}^{(8)}$ & $7.41 \cdot 10^{23}$ & $3.57 \cdot 10^{21}$ & $1.34 \cdot 10^{19}$ & $4.50 \cdot 10^{16}$ & $1.43 \cdot 10^{14}$ & \\
$D_{5,n}^{(8)}$ & $2.56 \cdot 10^{28}$ & $1.44 \cdot 10^{27}$ & $6.27 \cdot 10^{25}$ & $2.46 \cdot 10^{24}$ & $9.13 \cdot 10^{22}$ & \\
$D_{6,n}^{(8)}$ & $2.58 \cdot 10^{28}$ & $4.89 \cdot 10^{27}$ & $7.19 \cdot 10^{26}$ & $9.53 \cdot 10^{25}$ & $1.19 \cdot 10^{25}$ & \\
$D_{7,n}^{(8)}$ & $1.59 \cdot 10^{28}$ & $1.22 \cdot 10^{28}$ & $7.22 \cdot 10^{27}$ & $3.86 \cdot 10^{27}$ & $1.95 \cdot 10^{27}$ & \\
\bottomrule
\end{tabular}
\end{center}

\begin{center}
\begin{tabular}{l*{8}{c}}
\toprule
 & $n=6$ & $n=7$ & $n=8$ & $n=9$ & \\
\midrule
$D_{1,n}^{(8)} $ & $1.63 \cdot 10^{4}$ & $3.13$ & $5.91 \cdot 10^{-4}$ & $1.10 \cdot 10^{-7}$ & \\
$D_{2,n}^{(8)}$ & $4.35 \cdot 10^{8}$ & $2.92 \cdot 10^{5}$ & $1.93 \cdot 10^{2}$ & $1.25 \cdot 10^{-1}$ & \\
$D_{3,n}^{(8)}$ & $2.61 \cdot 10^{10}$ & $3.35 \cdot 10^{7}$ & $4.21 \cdot 10^{4}$ & $5.23 \cdot 10$ & \\
$D_{4,n}^{(8)}$ & $4.38 \cdot 10^{11}$ & $1.30 \cdot 10^{9}$ & $3.81 \cdot 10^{6}$ & $1.10 \cdot 10^{4}$ & \\
$D_{5,n}^{(8)}$ & $3.26 \cdot 10^{21}$ & $1.13 \cdot 10^{20}$ & $3.86 \cdot 10^{18}$ & $1.30 \cdot 10^{17}$ & \\
$D_{6,n}^{(8)}$ & $1.43 \cdot 10^{24}$ & $1.68 \cdot 10^{23}$ & $1.93 \cdot 10^{22}$ & $2.18 \cdot 10^{21}$ & \\
$D_{7,n}^{(8)}$ & $9.46 \cdot 10^{26}$ & $4.47 \cdot 10^{26}$ & $2.07 \cdot 10^{26}$ & $9.48 \cdot 10^{25}$ & \\
\bottomrule
\end{tabular}
\end{center}
}

\begin{multicols}{2}

 \end{multicols}
\end{document}